\documentclass[conference]{IEEEtran}
\IEEEoverridecommandlockouts

\usepackage{amsmath,amsfonts}
\usepackage{algorithmic}
\usepackage{algorithm}
\usepackage{array}
\usepackage[caption=false,font=normalsize,labelfont=sf,textfont=sf]{subfig}
\usepackage{textcomp}
\usepackage{stfloats}
\usepackage{url}
\usepackage{verbatim}
\usepackage{graphicx}
\usepackage{cite}
\usepackage{makecell}
\usepackage{subfig}  
\usepackage{amssymb}
\usepackage{mathrsfs}
\usepackage{amsthm}
\RequirePackage{epsf}
\usepackage{epstopdf}
\usepackage[printonlyused]{acronym}

\acrodef{C-V2X}{cellular vehicle-to-everything}
\acrodef{V2I}{vehicle-to-infrastructure}
\acrodef{V2V}{vehicle-to-vehicle}
\acrodef{V2P}{vehicle-to-pedestrian}
\acrodef{RB}{resource block}
\acrodef{NR}{new radio}
\acrodef{QoS}{quality-of-service}
\acrodef{SINR}{signal-to-interference-plus-noise ratio}
\acrodef{CSI}{channel state information}
\acrodef{BS}{base station}
\acrodef{TxV}{transmitting vehicle}
\acrodef{RxV}{receiving vehicle}
\acrodef{RSSI}{received signal strength indicator}
\acrodef{RSS}{received signal strength}
\acrodef{PDF}{probability distribution function}
\acrodef{RV}{random variable}
\acrodef{MSE}{mean square error}
\acrodef{HR}{hazard rate}
\acrodef{ITS}{intelligent transportation systems}
\acrodef{PSFCH}{physical sidelink feedback channel}
\acrodef{PUCCH}{physical uplink control channel}
\acrodef{RV}{random variable}
\acrodef{i.i.d.}{independent, identically distributed}
\acrodef{GMM}{Gaussian mixture model}
\acrodef{TVaR}{tail value at risk}
\acrodef{CRF}{conditional relative frequency}
\acrodef{ORF}{overall relative frequency}

\newtheorem{theorem}{Theorem}
\newtheorem{corollary}{Corollary}
\newtheorem{lemma}{Lemma}

\usepackage{siunitx}
\begin{document}

\title{A Resilience Perspective on C-V2X Communication Networks under Imperfect CSI \vspace{-0.37cm}}

\author{\normalsize Tingyu Shui\textsuperscript{1}, Walid Saad\textsuperscript{1}, and Mingzhe Chen\textsuperscript{2} \\ 
\textsuperscript{1}Bradley Department of Electrical and Computer Engineering, Virginia Tech, Arlington, VA, 22203, USA.\\
\textsuperscript{2}Department of Electrical and Computer Engineering and Frost Institute for Data Science and Computing, \\ University of Miami, Coral Gables, FL, 33146, USA. \\
Emails:\{tygrady, walids\}@vt.edu, mingzhe.chen@miami.edu\vspace{-0.7cm}}
\maketitle
\begin{abstract}
\Ac{C-V2X} networks provide a promising solution to improve road safety and traffic efficiency. One key challenge in such systems lies in meeting different \ac{QoS} requirements of coexisting vehicular communication links, particularly under imperfect \ac{CSI} conditions caused by the highly dynamic environment.
In this paper, a novel analytical framework for examining the resilience of \ac{C-V2X} networks in face of imperfect \ac{CSI} is proposed. In this framework, the adaptation phase of the \ac{C-V2X} network is studied, in which an adaptation power scheme is employed and the \ac{PDF} of the imperfect \ac{CSI} is estimated. Then, the resilience of \ac{C-V2X} networks is studied through two principal dimensions: remediation capability and adaptation performance, both of which are defined, quantified, and analyzed for the first time. Particularly, an upper bound on the estimation's \ac{MSE} is explicitly derived to capture the \ac{C-V2X}'s remediation capability, and a novel metric named \textit{\ac{HR}} is exploited to evaluate the \ac{C-V2X}'s adaptation performance. Afterwards, the impact of the adaptation power scheme on the \ac{C-V2X}'s resilience is examined, revealing a tradeoff between the \ac{C-V2X}'s remediation capability and adaptation performance.
Simulation results validate the framework's superiority in capturing the interplay between adaptation and remediation, as well as the effectiveness of the two proposed metrics in guiding the design of the adaptation power scheme to enhance the system's resilience.

\end{abstract}

\IEEEpeerreviewmaketitle

\acresetall
\vspace{-10pt}
\section{Introduction}
\vspace{-5pt}
\Ac{C-V2X} networks are expected to be a key enabler of \ac{ITS}. 3GPP Release 16 \cite{9345798} launched 5G \ac{NR} \ac{C-V2X} technology to enable heterogeneous vehicular communications using pervasive cellular \acp{BS}. 
To meet the diverse \ac{QoS} requirements of vehicles in \ac{C-V2X} networks, resource management based on real-time \ac{CSI} is necessary \cite{9530506}. However, managing network resources, such as power and spectrum, in \ac{C-V2X} networks is challenging since the imperfect \ac{CSI} caused by the dynamic vehicle network environment, e.g., vehicle mobility, can jeopardize the effectiveness of the resource management thereby degrading the \ac{C-V2X} network \ac{QoS}.

To address this degradation in \ac{C-V2X} network \ac{QoS}, there is a need for a resource management framework that instills \emph{resilience} to the system. As an extended concept of reliability and robustness, resilience represents \textit{``the capability of a system to adapt to unseen disruptions without prior information and finally remediate itself"}. Two key phases of resilience are \emph{adaptation} and \emph{remediation}. ``Adaptation" is the system's immediate reaction to the unseen disruption, which aims to maintain the operation of the system. Adaptation may last for a period during which the system must learn the disruption from its effect. Adaptation is followed by ``remediation", which represents the system's strategy to mitigate the disruption's impact in a long run, through the learned knowledge. 

A number of recent works \cite{7913583, 9400748, 8993812, 9374090, 9382930, 9857930, 10238756, 10213228} addressing resource management with imperfect \ac{CSI} mainly focused on reliability or robustness. The works in \cite{7913583, 9400748, 8993812, 9374090} studied the problem of power allocation and spectrum sharing for coexisting \ac{V2V} and \ac{V2I} links whereby the \ac{CSI} imperfection was assumed to follow a known distribution \cite{7913583, 9400748, 8993812} or to be bounded by a known range \cite{9374090}. However, the deployment of these model-based approaches in \cite{7913583, 9400748, 8993812, 9374090} may result in suboptimal performance in practice, as accurately modeling \ac{CSI} imperfections is inherently challenging in highly dynamic vehicular networks.
To address this issue, the works in \cite{9382930, 9857930, 10238756, 10213228} proposed the use of so-called \emph{data-driven approaches} that required no prior assumptions on the imperfection in the \ac{CSI}. Particularly, those works used the observed imperfect \ac{CSI} samples and proposed robust designs in which resources are allocated to guarantee a certain worst-case \ac{QoS}. However, the robust designs proposed in \cite{9382930, 9857930, 10238756, 10213228} will often result in overly conservative \ac{QoS} performance in \ac{C-V2X} networks.  
This is because these designs used the \ac{CSI} samples for the sole purpose of ensuring the worst-case \ac{QoS}, which may be detrimental to the overall achieved \ac{QoS}. Furthermore, none of the prior works in \cite{7913583, 9400748, 8993812, 9374090, 9382930, 9857930, 10238756, 10213228} considered the resilience of \ac{C-V2X} networks under imperfect \ac{CSI}. In particular, the \ac{QoS} performance of \ac{C-V2X} networks during the sample collection process, which corresponds to the adaptation phase of the system, was ignored. More importantly, prior works primarily focused on exploiting the imperfect \ac{CSI} samples, overlooking the \ac{C-V2X} networks' ability to actively explore more favorable samples to enhance network \ac{QoS}.




The main contribution of this paper is a novel analytical framework that defines, quantifies, and examines the resilience of a \ac{C-V2X} network in face of arbitrary unknown imperfection in the \ac{CSI}. Specifically, we consider the adaptation phase of the \ac{C-V2X} network, in which an adaptation power scheme is applied and the \ac{PDF} of the imperfection in \ac{CSI} is estimated. Given that the accuracy of the estimated \ac{PDF} is critical to effective resource management in the subsequent remediation phase, an upper bound on the \ac{MSE} of the estimated \ac{PDF} is derived and formally defined as the \emph{remediation capability} of the \ac{C-V2X} network. Moreover, the derived upper bound shows a tradeoff between remediation and adaptation. In particular, it demonstrates that achieving higher accuracy in estimating \ac{CSI} imperfections, i.e., a better remediation capability, will jeopardize the \ac{C-V2X}'s \ac{QoS} performance during the adaptation phase, and vice versa. Consequently, we leverage a novel metric named \textit{\ac{HR}} to evaluate the severity of \ac{QoS} degradations. Here, \ac{HR} provides a new perspective on the interplay between remediation capability and adaptation performance. To our best knowledge, \textit{this is the first work that analyzes the resilience of \ac{C-V2X} in face of imperfect \ac{CSI} from the perspective of adaption and remediation}. Simulation results validate the framework's superiority in capturing the interplay between adaptation and remediation, as well as the effectiveness of the two proposed metrics in guiding the design of adaptation power scheme to enhance system's resilience.

\vspace{-5pt}
\section{System Model}
\vspace{-3pt}
\subsection{\ac{C-V2X} Network Model}
\begin{figure}[t]
	\centering
	\includegraphics[scale=0.46]{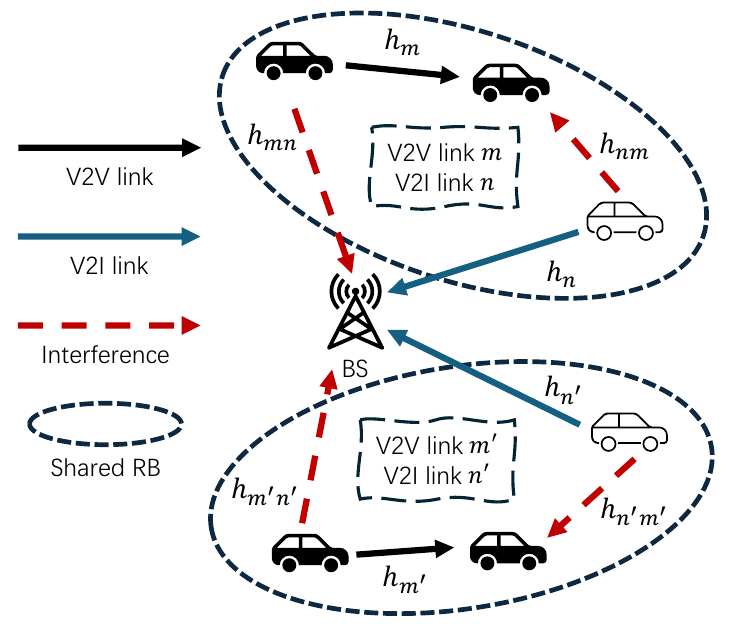}
 \vspace{-10pt}
	\caption{\small{System model of the considered \ac{C-V2X} network.}}
 \vspace{-10pt}
 \label{System Model}
  \vspace{-10pt}
\end{figure}

\vspace{-3pt}
Consider a \ac{C-V2X} network supporting multiple \ac{V2I} links through the Uu interface designed for cellular uplink and downlink transmission \cite{9345798}. A centralized \ac{BS} allocates orthogonal \acp{RB} to each \ac{V2I} link to avoid mutual interference. Within the coverage of the \ac{BS}, multiple \ac{V2V} links that use transmission mode-1 through \ac{NR} sidelinks are deployed. The \ac{V2V} links will reuse the \ac{RB}s allocated to the \ac{V2I} links to transmit time-sensitive and safety-critical messages, as shown in Fig. \ref{System Model}. Since the uplink transmission of \ac{V2I} links are generally less used, the \ac{RB}s are reused during the uplink. Moreover, we assume that a certain \ac{V2V} link will only reuse a single \ac{RB} and the allocated \ac{RB} of a certain \ac{V2I} link can be only shared with one \ac{V2V} link. Under \ac{NR} \ac{C-V2X} transmission mode 1, the \ac{BS} will determine the pairing of \ac{V2V} links and \ac{V2I} links sharing the same \ac{RB} and allocate the transmit power of these links according to the \ac{CSI}. The optimal pairing can be obtained by using the Hungarian method \cite{9382930} after the power allocation of an arbitrary pair is derived. Thus, hereinafter, we only focus on an arbitrary pair.

\vspace{-4pt}
\subsection{\ac{QoS} of the \ac{V2V} and \ac{V2I} links}
\vspace{-4pt}
Without loss of generality, we assume that \ac{V2I} link $n$ is sharing its \ac{RB} with \ac{V2V} link $m$. The transmit power of \ac{V2I} link $n$ and \ac{V2V} link $m$ are $P_n$ and $P_m$ respectively.
Let $h_n$ be the channel gain on the \ac{V2I} link $n$ and $h_{mn}$ be the channel gain on the interference link from \ac{V2V} link $m$ to \ac{V2I} link $n$, the \ac{SINR} over the \ac{V2I} link $n$ can be obtained as $\gamma_n= \frac{ P_{n} h_n }{ P_{m} h_{mn} + \sigma^2 }$, where $\sigma^2$ is the power of additive white Gaussian noise. Similarly, we can obtain the \ac{SINR} over the \ac{V2V} link $m$ as $\gamma_m= \frac{ P_{m} h_m }{ P_{n} h_{nm} + \sigma^2 }$, where $h_m$ and $h_{nm}$ represent the channel gain on \ac{V2V} link $m$ and the interference link from \ac{V2I} link $n$ to \ac{V2V} link $m$. Here, the channel gain $h \in \mathcal{H} = \left\{ h_n, h_m, h_{nm}, h_{mn} \right\}$ is modeled as $h = L | g | ^2$ where $L$ is the large-scale fading and $g$ is the small-scale fading. The large-scale fading is further modeled as $L = G \zeta d^{-\alpha}$ with path loss gain $G$, log-normal shadow fading gain $\zeta$, path loss exponent $\alpha$ and the link distance $d$ of the vehicular link. Assuming a Rayleigh channel model\cite{7913583}, the small-scale fading is represented by $g \sim \mathcal{C N}(0,1)$. Given the different use cases of vehicular communication links, the heterogeneous \ac{QoS} requirements on \ac{V2I} link $n$ and \ac{V2V} link $m$ will be given by: 
\vspace{-5pt}
\begin{equation}
\label{throughput}
    R_m \triangleq B \log (1+\gamma_n) \geq R_0,
\end{equation}
\vspace{-15pt}
\begin{equation}
\label{delay}
   \tau_n \triangleq \frac{D}{B \log (1+\gamma_m)} \leq \tau_0,
   \vspace{-5pt}
\end{equation}
where $B$ is the bandwidth of each \ac{RB}, $D$ is the \ac{V2V} link packet size, and $R_0$ and $\tau_0$ are respectively the given throughput and delay requirements. 

The power allocation problem, widely explored in recent literature \cite{7913583, 9400748, 8993812, 9374090, 9382930, 9857930, 10238756, 10213228}, seeks to determine the values of $P_m$ and $P_n$ that meets the \ac{QoS} requirements in \eqref{throughput} and \eqref{delay}. The difficulty of the problem lies in the fact that obtaining an accurate \ac{CSI} in $\mathcal{H}$ can be challenging in highly dynamic \ac{C-V2X} networks. Specifically, in \ac{NR} \ac{C-V2X} mode 1, all involved vehicular links are required to report their \ac{CSI} to the \ac{BS}, either directly through \ac{PUCCH} or relayed by the \ac{PSFCH} \cite{8998153}. Thus, the value for $P_m$ and $P_n$ can be only determined according to the imperfect \ac{CSI} $\hat{h} \in \mathcal{\hat{H}} = \left\{ \hat{h}_n, \hat{h}_m, \hat{h}_{nm}, \hat{h}_{mn} \right\}$, which may not satisfy \eqref{throughput} and \eqref{delay}.

\vspace{-5pt}
\subsection{Imperfect \ac{CSI} model}
\vspace{-5pt}
We assume that the large-scale fading can be perfectly estimated by the \ac{BS}, since $L$ is dominantly determined by the locations of the vehicles and varies on a slow scale.
Therefore, we only consider the imperfect small-scale fading. We assume $\hat{g}_n = g_n$ and $\hat{g}_{mn} = g_{mn}$, while $\hat{g}_m \neq g_m$ and $\hat{g}_{nm} \neq g_{nm}$. The reason for assuming $\hat{g}_n = g_n$ and $\hat{g}_{mn} = g_{mn}$ is that $\hat{g}_n$ and $\hat{g}_{mn}$ are directly estimated by the \ac{BS}. However, $\hat{g}_m$ and $\hat{g}_{nm}$ are both relayed to the \ac{BS} through \ac{PUCCH} and \ac{PSFCH}, which results in severe \ac{CSI} feedback delay.  Moreover, the Doppler shift caused by the relative movement between vehicles should be considered when modeling $\hat{g}_m$ and $\hat{g}_{nm}$. For $\hat{g}_m$, since \ac{V2V} link are generally established when the relative movement of two vehicles are comparatively stable and predictable \cite{8993812}, we directly model $\hat{g}_m$ through the first-order Gauss-Markov process:
\vspace{-7pt}
\begin{equation}
\label{i v2v}
   | g_{m} | ^2 = | \hat{g}_{m} |^2 + (1 - \delta_m^2) \left( | {e}_{m} |^2 - | \hat{g}_{m} |^2 \right),
   \vspace{-3pt}
\end{equation}
where $\delta_m = J_0\left(2 \pi f_D \Delta T\right)$ is the coefficient given by Jakes statistical model \cite{7913583} and ${e}_{m} \sim \mathcal{C N}(0,1)$ is the error term. Specifically, $J_0$ is the zero-order Bessel function of the first kind, $\Delta T$ represents the \ac{CSI} feedback delay, and $f_D = \frac{v f_c}{c}$ is the maximum Doppler frequency with $c$ being the speed of light, where $v$ and $f_c$ are the vehicle speed and carrier frequency respectively. Generally, $\delta_m$ can be derived by the vehicles in \ac{V2V} link $m$ \cite{9400748, 8993812}, because the \ac{CSI} feedback delay $\Delta T$ can be obtained by adding time stamp when estimating $\hat{g}_{m}$.
However, it is almost impossible to find a suitable model for $\hat{g}_{nm}$ of the interference link. This is because the relative movement between the \ac{V2V} link $m$ and \ac{V2I} link $n$ is highly dynamic and hard to model. Furthermore, the dynamic pairing of \ac{V2V} link $m$ and \ac{V2I} link $n$ makes it impractical to assume prior information on $\hat{g}_{nm}$. As a result, we only assume an additive error ${e}_{nm}$, where ${e}_{nm} \sim \mathcal{E}$ is a \ac{RV} that follows an unknown distribution $\mathcal{E}$. Then, the model for $\hat{g}_{nm}$ is given by:
\vspace{-7pt}
\begin{equation}
\label{i v2i}
   | g_{nm} | ^2 = | \hat{g}_{nm} |^2 + {e}_{nm}.
   \vspace{-3pt}
\end{equation}

Due to the imperfect \ac{CSI} in \eqref{i v2v} and \eqref{i v2i}, we can only seek to satisfy \eqref{delay} with a high probability $P_0$, which is given by:
\vspace{-7pt}
\begin{equation}
\label{prob}
   \mathbb{P} \left\{\frac{P_m L_m}{ \gamma_0 } (1 - \delta_{m}^2) | {e}_{m} |^2  - P_n L_n {e}_{nm} \geq b \right\} \geq P_0,
   \vspace{-5pt}
\end{equation}
where $b = \sigma^2 + P_n L_{nm} | \hat{g}_{nm} | ^2 - \frac{P_m L_m \delta_m^2 | \hat{g}_{m} | ^2}{\gamma_0}$ is determined by the imperfect \ac{CSI} $\hat{g}_{m}$ and $\hat{g}_{nm}$ and $\gamma_0 = 2^{\frac{D}{B\tau_0}}-1$ is a constant. However, computing the probability in \eqref{prob} is still challenging because the \ac{PDF} of distribution $\mathcal{E}$ is unknown. Moreover, the distribution $\mathcal{E}$ may be time-varying due to the highly dynamic environment, which exacerbates this problem.
\vspace{-15pt}
\section{Resilience Analysis}
\vspace{-5pt}
To meet the \ac{QoS} requirement in \eqref{prob}, in this section, we propose a novel framework for instilling resilience into the \ac{C-V2X} network. Particularly, the \ac{PDF} of distribution $\mathcal{E}$ is firstly estimated by the \ac{BS}, through a dedicated \emph{adaptation phase}. Afterwards, the effectiveness of the adaptation is captured by the \ac{MSE} of the estimation, whose upper bound is derived and defined as the \emph{remediation capability} of the \ac{C-V2X} network. A tradeoff between the \ac{QoS} during the adaptation phase and the remediation capability is then explored. Finally, a novel metric named \emph{hazard rate} is incorporated and analyzed to capture the \ac{C-V2X}'s adaptation performance.
\vspace{-4pt}
\subsection{Deconvolution Estimation during Adaptation Phase}
\vspace{-4pt}
To estimate the \ac{PDF} of $\mathcal{E}$, the \ac{BS} will switch to a dedicated adaptation phase lasting for $T$ time slots, where an adaptation power profile consisting of $P_n^{\textrm{a}}$ on the \ac{V2I} link $n$ and $P_m^{\textrm{a}}$ on \ac{V2V} link $m$, is used. Specifically, $P_n^{\textrm{a}}$ and $P_m^{\textrm{a}}$ aim not only to meet the \ac{QoS} requirements of the \ac{C-V2X} network, but also to estimate the \ac{PDF} of the distribution $\mathcal{E}$. After estimating the \ac{PDF}, the \ac{BS} will switch to a remediation phase, where \eqref{prob} is derived for optimizing $P_n$ and $P_m$ based on imperfect \ac{CSI}.\footnote{To enhance the resilience of the \ac{C-V2X} network, this power allocation optimization in remediation phase is equally as important as the design of adaptation power scheme in adaptation phase. However, we leave this problem for future work due to space limitation.} During the adaptation phase, the large-scale fading parameters $L^{\textrm{a}}_m$ and $L^{\textrm{a}}_{nm}$ are assumed to be invariant. Then, we can leverage the \ac{RSS} at the receiving vehicle of \ac{V2V} link $m$ to estimate the \ac{PDF} of $\mathcal{E}$. Particularly, the \ac{RSS} at the receiving vehicle of \ac{V2V} link $m$ at time slot $t$ can be given as ${r}_t = P_n^{\textrm{a}} L_{nm}^{\textrm{a}} | g_{nm,t} | ^2 + P_m^{\textrm{a}} L_{m}^{\textrm{a}} | g_{m,t} | ^2 + \sigma^2$. The \ac{RSS} ${r}_t$ will be fed back to the \ac{BS} through \ac{PUCCH} and \ac{PSFCH}, forming an \ac{RSS} set ${\mathcal{R}} = \left\{ {r}_1, \ldots, {r}_T \right\}$. Correspondingly, the \ac{BS} can derive the ideal \ac{RSS} at time slot $t$ based on $ \hat{g}_{m,t}$ and $\hat{g}_{nm,t}$. This ideal \ac{RSS} is given by $\hat{r}_t = P_n^{\textrm{a}} L_{nm}^{\textrm{a}} | \hat{g}_{nm,t} | ^2 + P_m^{\textrm{a}} L_{m}^{\textrm{a}} | \hat{g}_{m,t} | ^2 + \sigma^2$. Thus, the \ac{BS} can form an ideal \ac{RSS} set $\hat{\mathcal{R}} = \left\{ \hat{r}_1, \ldots, \hat{r}_T \right\}$ at the end of the adaption phase. To this end, the \ac{BS} can collect a sequence of data samples $\mathcal{Z} = \left\{ z_1, \ldots, z_T \right\}$, where $z_t$ is defined as
\vspace{-7pt}
\begin{equation}
   z_t \triangleq \frac{{r}_t - \hat{r}_t}{P_n^{\textrm{a}} L_{nm}^{\textrm{a}}} + \frac{P_m^{\textrm{a}} L_{m}^{\textrm{a}} }{P_n^{\textrm{a}} L_{nm}^{\textrm{a}} }(1 - \delta_{m}^2)| \hat{g}_{m,t} | ^2.
      \vspace{-3pt}
\end{equation}
From \eqref{i v2v} and \eqref{i v2i}, we have:
\vspace{-7pt}
\begin{equation}
   z_t = {e}_{nm, t} + \frac{P_m^{\textrm{a}} L_{m}^{\textrm{a}} }{P_n^{\textrm{a}} L_{nm}^{\textrm{a}} }(1 - \delta_{m}^2) | {e}_{m, t} |^2,
   \vspace{-5pt}
\end{equation}
where ${e}_{nm, t}$ and ${e}_{m, t}$ are the realizations of the error terms in \eqref{i v2v} and \eqref{i v2i} at time slot $t$. Note that neither ${e}_{nm, t}$ nor ${e}_{m, t}$ can be directly obtained by the \ac{BS}; however, the value of $z_t$ is accessible. Clearly, $\mathcal{Z}$ is essentially a sequence of \ac{i.i.d.} samples from \ac{RV} $Z$ given by:
\vspace{-7pt}
\begin{equation}
\label{deconv}
   Z = {e}_{nm} + Y,
   \vspace{-5pt}
\end{equation}
where ${e}_{nm} \sim \mathcal{E}$ and $Y = \frac{P_m^{\textrm{a}} L_{m}^{\textrm{a}} }{P_n^{\textrm{a}} L_{nm}^{\textrm{a}} } (1 - \delta_{m}^2) | {e}_{m} |^2$.

Our goal is to derive the \ac{PDF} of the error distribution $\mathcal{E}$ through $\mathcal{Z}$. As shown in \eqref{deconv}, $Z$ is the sum of two independent \ac{RV} ${e}_{nm}$ and $Y$, where $Y$ follows an exponential distribution because ${e}_{m} \sim \mathcal{C N}(0,1)$. Deriving the \ac{PDF} of $\mathcal{E}$ through $\mathcal{Z}$ is essentially a deconvolution problem \cite{10.3150/08-BEJ146}. For notation simplicity, we define $f_E$ as the \ac{PDF} of distribution $\mathcal{E}$. Given the Fourier transforms of $f_Z$, $f_E$, and $f_Y$ denoted by $\mathcal{F}\left\{ f_Z \right\}$, $\mathcal{F}\left\{ f_E \right\}$, $\mathcal{F}\left\{ f_Y \right\}$, we can apply the Pascal theorem and approximate $\mathcal{F}\left\{ f_E \right\}$ \cite{doi:10.1080/02331889008802238} as follows:
\vspace{-7pt}
\begin{equation}
\begin{aligned}
\label{Fourier}
    F\left\{ f_E \right\} 
    = & \frac{F\left\{ f_Z \right\}}{F\left\{ f_Y \right\}} \approx \frac{1}{T}\sum_{t=1}^T e^{-jwz_t}(1+\frac{jw}{\lambda_Y}),
\end{aligned}
       \vspace{-5pt}
\end{equation}
where $F\left\{ f_Z \right\}$ is approximated by its empirical counterpart $\frac{1}{T}\sum_t^T e^{-jwz_t}$ from $\mathcal{Z}$ and $F\left\{ f_Y \right\}$ is obtained by $Y \sim \text{exp} (\lambda_Y)$ with $\lambda_Y = \frac{P_n^{\textrm{a}} L_{nm}^{\textrm{a}} }{P_m^{\textrm{a}} L^{\textrm{a}}_{m} (1 - \delta_{m}^2)}$.
From \eqref{Fourier}, we can derive $\hat{f}_E$, the estimation of $f_E$, by the inverse Fourier transform:
\vspace{-7pt}
\begin{equation}
\begin{aligned}
\label{hat pdf}
    \hat{f}_E 
    = & \frac{1}{2 \pi T}  \sum_{t=1}^T \int_{-\infty}^{\infty}  e^{-jw(z_t-e_{nm})}  (1+\frac{jw}{\lambda_Y}) dw \\
    \approx & \frac{1}{2 \pi T}  \sum_{t=1}^T \int_{-K\pi}^{K \pi}  e^{-jw(z_t-e_{nm})}  (1+\frac{jw}{\lambda_Y}) dw,
\end{aligned}
       \vspace{-5pt}
\end{equation}
where the integral is truncated into $[-K \pi, K \pi]$ with constant $K$ to ensure the convergence of $\hat{f}_E$ \cite{10.3150/08-BEJ146}.
\vspace{-4pt}
\subsection{Remediation Capability Analysis}
\vspace{-4pt}
With the estimator $\hat{f}_E$ in \eqref{hat pdf}, the \ac{BS} can compute the probability in \eqref{prob} given the imperfect \ac{CSI} $\mathcal{\hat{H}}$. Clearly, the accuracy of the estimator is critical for meeting the \ac{QoS} requirement in \eqref{prob} when we optimize $P_n$ and $P_m$ during remediation. Therefore, we define the \ac{MSE} of the estimator $\hat{f}_E$, i.e., $\mathbb{E}\left[  (f_{E} - \hat{f}_{E})^2 \right]$, as the \ac{C-V2X}'s remediation capability. Next, we derive an upper bound of the \ac{MSE}.
\vspace{-3pt}
\begin{theorem}
\label{theorem 1}
The remediation capability of the \ac{C-V2X} network, i.e., the \ac{MSE} of the estimator $\hat{f}_E$, is upper bounded as:
\vspace{-10pt}
\begin{equation}
\label{bound}
    \begin{aligned}
        & \mathbb{E}\left[  (f_{E} - \hat{f}_{E})^2 \right] 
        \leq  \frac{1}{4\pi^2} \left( \int_{w\geq |K\pi|} e^{jwe_{nm}} F\left\{ f_E \right\} dw \right)^2 \\
        & + \frac{K^2}{4T}\left[ \sqrt{1+ o^2} + \frac{\ln{\left(o + \sqrt{1+ o^2}\right)}}{o} \right]^2,
    \end{aligned}
    \vspace{-10pt}
\end{equation}
where $o = K \pi (1-\delta_m^2) \frac{P^{\textrm{a}}_m L^{\textrm{a}}_m }{P^{\textrm{a}}_n L^{\textrm{a}}_n}$.
\vspace{-5pt}
\begin{proof}
See Appendix A.
\end{proof}
\end{theorem}
\vspace{-5pt}
\addtolength{\topmargin}{0.05in}
Theorem \ref{theorem 1} shows that the remediation capability of the \ac{C-V2X} network is upper bounded by the sum of two terms: the first one related to the unknown error distribution $\mathcal{E}$ and the second one as a function of the adaptation power scheme $P^{\textrm{a}}_m$ and $P^{\textrm{a}}_n$. Moreover, from \eqref{bound}, we can observe that the second term is monotonously increasing with $o = K \pi (1-\delta_m^2) \frac{P^{\textrm{a}}_m L^{\textrm{a}}_m }{P^{\textrm{a}}_n L^{\textrm{a}}_n}$. Thus, a high power $P_n^{\textrm{a}}$ on the \ac{V2I} link can enhance the \ac{C-V2X}'s remediation capability by obtaining an accurate estimation on $f_{E}$. The reason is that the error $e_{nm}$ becomes the dominant component in $Z$ when a high power $P_n^{\textrm{a}}$ is applied, as shown in \eqref{deconv}. Conversely, employing a high power $P_m^{\textrm{a}}$ on the \ac{V2V} link will compromise the \ac{C-V2X}'s remediation capability, since $Y$ becomes the dominant component in $Z$ other than $e_{nm}$, which, in turn, decreases the accuracy of the estimation $\hat{f}_{E}$. Moreover, we can see that, as the adaptation phase lasts longer, i.e., a higher $T$ is allowed, the system's remediation capability can be improved.

More insights about the design of the adaptation power scheme can be drawn from Theorem \ref{theorem 1}. First, in order to enhance the remediation capability, the \ac{BS} should properly pair the \ac{V2I} link and \ac{V2V} link according to the large-scale fading parameters $L_m^{\textrm{a}}$ and $L_n^{\textrm{a}}$ and the coefficient $\delta_m$ before switching to the adaptation phase.
Moreover, the design of the adaptation power scheme $P^{\textrm{a}}_m$ and $P^{\textrm{a}}_n$ exhibits a tradeoff between the \ac{C-V2X}'s remediation capability and its \ac{QoS} performance during the adaptation phase. In particular, we may simply implement the minimal $P_m^{\textrm{a}}$ and the maximal $P_n^{\textrm{a}}$ to obtain an accurate estimation $\hat{f}_{E}$, which, however, will jeopardize the \ac{QoS} performance of the \ac{V2V} link $m$. 
\vspace{-4pt}
\subsection{Adaptation Performance Analysis}
\vspace{-4pt}
From a resilience perspective, the \ac{C-V2X} network is expected to achieve a high remediation capability, on the condition that the \ac{QoS} during adaptation phase is not significantly compromised. Current works \cite{7913583, 9400748, 8993812, 9374090, 9382930, 9857930, 10238756, 10213228} basically focus on the reliable or robust design of \ac{C-V2X} network under imperfect \ac{CSI}. Thus, metrics emphasizing the \ac{C-V2X} network's capability to strictly satisfy \ac{QoS} requirements are used, e.g., the \ac{QoS} outage probability or the worst-case \ac{QoS}. However, according to the definition of resilience \cite{STERBENZ20101245}, the adaptation phase entails more than merely satisfying \ac{QoS} requirements. Specifically, the \ac{BS} should react to the imperfect \ac{CSI} to learn its impact as well during the adaptation phase, where the learned knowledge will be utilized during the following remediation phase. 

Thus, we need a new metric to capture the adaptation performance and demonstrate the interplay between adaptation and remediation. On the one hand, short-term \ac{QoS} degradation during adaptation can be manageable in case it allows achieving a high remediation capability in the long term. On the other hand, this short-term \ac{QoS} degradation should be mild to avoid dire consequences. For example, if the V2V delay requirement is $\tau_0 = 10$~ms, the case in which the system is affected by a QoS degradation that keeps the delay within $[10,20]$~ms would be preferable than one in which the delay fluctuates in the range $[30,40]$~ms.
From this perspective, we adopt the concept of \emph{hazard rate} \cite{Brody2007} to evaluate the system's \ac{QoS} during the adaptation phase.
Formally, given the different \ac{QoS} requirements $R_0$ and $\tau_0$ on the \ac{V2I} and \ac{V2V} links, the \ac{HR}s on the two links are respectively defined as: 
\vspace{-7pt}
\begin{equation}
\label{hr v2i}
    \Lambda^{-}_I(R_0) \triangleq \lim_{\Delta r \rightarrow 0} \frac{\mathbb{P} \left\{ R_0 - \Delta r \leq R \leq R_0 \right\}}{\Delta r \mathbb{P} \left\{ R \leq R_0 \right\}},
\end{equation}
\vspace{-10pt}
\begin{equation}
\label{hr v2v}
    \Lambda^{+}_V(\tau_0) \triangleq \lim_{\Delta \tau \rightarrow 0} \frac{\mathbb{P} \left\{ \tau_0 \leq \tau \leq \tau_0 + \Delta \tau \right\}}{\Delta \tau \mathbb{P} \left\{ \tau \geq \tau_0 \right\}}.
    \vspace{-3pt}
\end{equation}

Note that \eqref{hr v2i} and \eqref{hr v2v} consider the ergodic \ac{QoS} over the small-scale fading during adaptation.
According to the above definition, \ac{HR} quantifies the system’s capability to sustain \ac{QoS} close to the specified \ac{QoS} requirements, given that the \ac{QoS} requirement has already been violated. 
In other words, conditional on the \ac{QoS} requirement not being satisfied, a high \ac{HR} during adaptation  ensures a high probability of maintaining the \ac{QoS} near the specified requirement. Take $\tau_0 = 10$ ms for example, a high \ac{HR} $\Lambda^{+}_V(\tau_0)$ implies an increased likelihood of preserving $\tau$ close to $\tau_0 = 10$~ms when $\tau > \tau_0$. The explicit expression of the \ac{HR}s in \eqref{hr v2i} and \eqref{hr v2v} is derived next.
\begin{lemma}
\label{lemma 1}
The \ac{HR} $\Lambda^{+}_V(\tau_0)$ and $\Lambda^{-}_I(R_0)$ on \ac{V2V} link and \ac{V2I} link are respectively  given by
\vspace{-5pt}
\begin{equation}
\label{HR v2v}
\begin{aligned}
    \Lambda^{+}_V(\tau_0) = D_V e^{-\frac{\sigma^2\gamma_V}{P_m^{\textrm{a}}L_m^{\textrm{a}}}} \frac{ \frac{P_n^{\textrm{a}}L_{nm}^{\textrm{a}}}{P_m^{\textrm{a}}L_m^{\textrm{a}}} + \frac{\sigma^2}{P_m^{\textrm{a}} L_m^{\textrm{a}}} \left( 1 + \frac{P_n^{\textrm{a}}L_{nm}^{\textrm{a}}}{P_m^{\textrm{a}}L_m^{\textrm{a}}} \gamma_V \right)  }{\left( 1 + \frac{P_n^{\textrm{a}}L_{nm}^{\textrm{a}}}{P_m^{\textrm{a}}L_m^{\textrm{a}}} \gamma_V - e^{-\frac{\sigma^2\gamma_V}{P_m^{\textrm{a}}L_m^{\textrm{a}}}} \right)^2 },
\end{aligned}
\end{equation}
\vspace{-10pt}
\begin{equation}
\label{HR v2i}
\begin{aligned}
    \Lambda^{-}_I(R_0) = D_I e^{-\frac{\sigma^2\gamma_I}{P_n^{\textrm{a}}L_{nm}^{\textrm{a}}}} \frac{ \frac{P_m^{\textrm{a}}L_{m}^{\textrm{a}}}{P_n^{\textrm{a}}L_{nm}^{\textrm{a}}} + \frac{\sigma^2}{P_n^{\textrm{a}} L_{nm}^{\textrm{a}}} \left( 1 + \frac{P_m^{\textrm{a}}L_{m}^{\textrm{a}}}{P_n^{\textrm{a}}L_{nm}^{\textrm{a}}} \gamma_I \right) }{ \left( 1 + \frac{P_m^{\textrm{a}}L_{m}^{\textrm{a}}}{P_n^{\textrm{a}}L_{nm}^{\textrm{a}}} \gamma_I - e^{-\frac{\sigma^2\gamma_I}{P_n^{\textrm{a}}L_{nm}^{\textrm{a}}}} \right)^2 },
\end{aligned}
    \vspace{-5pt}
\end{equation}
where $D_V = \frac{\ln{2}D2^{\frac{D}{B\tau_0}}}{B\tau_0^2} $, $\gamma_V = 2^{\frac{D}{B\tau_0}} - 1$, $D_I = \frac{\ln{2}2^{\frac{R_0}{B}}}{B} $, and $\gamma_I = 2^{\frac{R_0}{B}} - 1$.
\end{lemma}
\vspace{-8pt}
\begin{proof}
The proof was omitted due to space limitation.
\end{proof}
\vspace{-5pt}

Aligned with resilience, \ac{HR} is not intended to strictly prevent \ac{QoS} violations, but rather to mitigate the severity of such violations. Thus, Lemma \ref{lemma 1} provides a guidance on the design of the adaptation power scheme $P_m^{\textrm{a}}$ and $P_n^{\textrm{a}}$. Precisely, a high \ac{HR} can enhance the \ac{C-V2X} network's remediation capability while keeping a mild \ac{QoS} degradation during adaptation.
To this end, we can define the weighted sum \ac{HR} as a metric to capture the adaptation performance of the \ac{C-V2X}:
\vspace{-7pt}
\begin{equation}
\label{HR}
    \Lambda = \lambda_V \Lambda^{+}_V(\tau_0) + \lambda_I  \Lambda^{-}_I(R_0),
    \vspace{-5pt}
\end{equation}
where $\lambda_V \geq 0$ and $\lambda_I \geq 0$ represent the priority of the \ac{V2I} and \ac{V2V} links and $\lambda_V + \lambda_I = 1$. With \eqref{bound} and \eqref{HR}, we can design the adaptation power scheme $P_n^{\textrm{a}}$ and $P_n^{\textrm{a}}$ to balance the tradeoff between the \ac{C-V2X} network's remediation capability and adaptation performance, which will be left for future work due to space limitation.
\vspace{-5pt}
\section{Simulation Results and Analysis}
\vspace{-5pt}
\begin{table}[t]
\centering
\caption{Path loss model}
\scriptsize
\vspace{-5pt}
\label{channel simu}
    \begin{tabular}{|c|c|c|}
    \hline
    \textbf{Channels}    & \textbf{Path loss}     & \textbf{\makecell[c]{Shadowing \\ standard deviation}}    \\ \hline
    $h_m$ & WINNER + B1 (LOS) \cite{kyosti2007winner} & $3$ dB  \\ \hline
    $h_n$ & $128.1 + 37.6 \log_{10}d_n$ ($d_n$ in km)  & $3$ dB  \\ \hline
    $h_{nm}$ & WINNER + B1 (NLOS)  \cite{kyosti2007winner}          & $4$ dB  \\ \hline
    $h_{mn}$ & WINNER + B1 (NLOS)           & $4$ dB  \\ \hline
    \end{tabular}
    \vspace{-0.5cm}
\end{table}
\begin{table}[t]
\centering
\caption{Adaptation power scheme}
\vspace{-5pt}
\scriptsize
\label{PA simu}
    \begin{tabular}{|c|c|c|}
    \hline
    \textbf{Power allocation schemes}    & \textbf{$P_n^{\textrm{a}}$}     & \textbf{$P_m^{\textrm{a}}$}    \\ \hline
    PA I & $\frac{1}{2} P_n^{\textrm{max}}$ & $\frac{1}{2} \min \left( P_n^{\textrm{max}}\frac{L_{nm}^{\textrm{a}}}{L_{m}^{\textrm{a}}}, P_m^{\textrm{max}}\right)$   \\ \hline
    PA II & $ P_n^{\textrm{max}}$  & $  \min \left( P_n^{\textrm{max}}\frac{L_{nm}^{\textrm{a}}}{L_{m}^{\textrm{a}}}, P_m^{\textrm{max}}\right)$   \\ \hline
    PA III & $ \frac{1}{2} P_n^{\textrm{max}}$  & $2 \min \left( P_n^{\textrm{max}}\frac{L_{nm}^{\textrm{a}}}{L_{m}^{\textrm{a}}}, P_m^{\textrm{max}}\right)$   \\ \hline
    \end{tabular}
    \vspace{-0.5cm}
\end{table}
For our simulations, we consider a \ac{C-V2X} network covering a circular area of radius $R = 200$ m, where the transmitting vehicles of both \ac{V2I} and \ac{V2V} links are randomly located according to a uniform distribution. The distance between transmitting vehicle and receiving vehicle of \ac{V2V} link $m$ is chosen randomly between $20$ m and $60$ m according to a uniform distribution. The noise spectrum density is $-174$ dBm/Hz. The other parameters are given by $B = 10$~MHz, $f_c = 5.9$~GHz, $\Delta T = 0.5$~ms, $v = 10$~m/s, $D=10,000$~bits, $\tau_0 = 10$~ms, $R_0 = 1$~Mbps, and $K = 10$. The path loss exponents for all channel models are $\alpha = 3$ and the path loss model is given in Table \ref{channel simu}. 
For our simulation, the distribution $\mathcal{E}$ of the error in imperfect \ac{CSI} is given as a \ac{GMM} with two component $x_1 \sim \mathcal{N}(0.2,0.04)$ and $x_2 \sim \mathcal{N}(0.8,0.02)$ with equal weight. This defined \ac{GMM} is only used for validating our method. All statistical results are averaged over \num{10000} channel realizations with $T =$ \num{1000}.


\begin{figure}[t]
\captionsetup[subfigure]{font=footnotesize}  
	\centering
	\subfloat[]
 {\includegraphics[width=0.49\columnwidth]{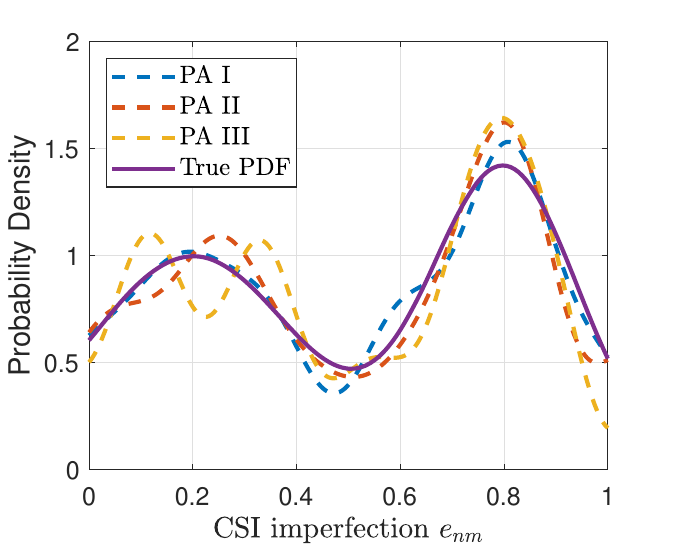}\label{pdf}}
 \hspace{1pt}
	\subfloat[]
 {\includegraphics[width=0.49\columnwidth]{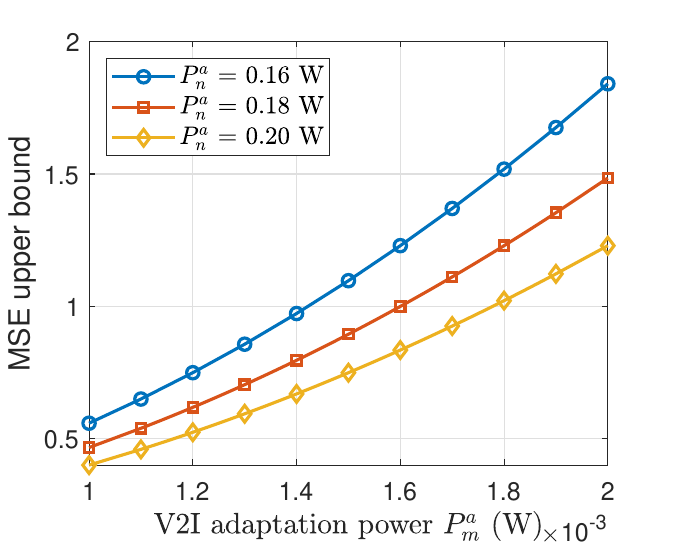}\label{mse}}
	\caption{\small{ Deconvolution-based estimation through the adaptation power scheme: a) Estimator $\hat{f}_E$ in \eqref{hat pdf}, b) Upper bound on the \ac{MSE}.}
 }
 \label{estimation}
  \vspace{-0.5cm}
\end{figure}
Fig. \ref{estimation} shows the impact of adaptation power scheme on the \ac{C-V2X}'s remediation capability. Table \ref{PA simu} lists three possible adaptation power schemes, PA I, PA II, and PA III, designed according to Theorem \ref{theorem 1} with $P_n^{\textrm{max}} = 23$ dBm and $P_m^{\textrm{max}} = 20$ dBm. Specifically, based on \eqref{bound}, the \ac{MSE} upper bound of PA I is equal to that of PA II, both of which are smaller than that of PA III. From Fig. \ref{pdf}, we can observe that the estimation under PA I and PA II are more accurate than that under PA III. In other words, PA I and PA II provide a higher remediation capability to the \ac{C-V2X} network. The reason is that the upper bound of the \ac{MSE} is monotonously increasing with the parameter $o$ and we can simply derive that the parameters $o_{\textrm{I}}$, $o_{\textrm{II}}$ and $o_{\textrm{III}}$ of PA I, PA II, and PA III have the relationship $o_{\textrm{I}} = o_{\textrm{II}} = \frac{1}{2} o_{\textrm{III}}$. Fig. \ref{mse} shows how the second term in \eqref{bound} is impacted by the adaptation power scheme. Clearly, a higher $P_m^{\textrm{a}}$ will threaten the estimation accuracy and lead to a decreased remediation capability.

\begin{figure}[t]
\captionsetup[subfigure]{font=footnotesize}  
	\centering
	\subfloat[]
 {\includegraphics[width=0.49\columnwidth]{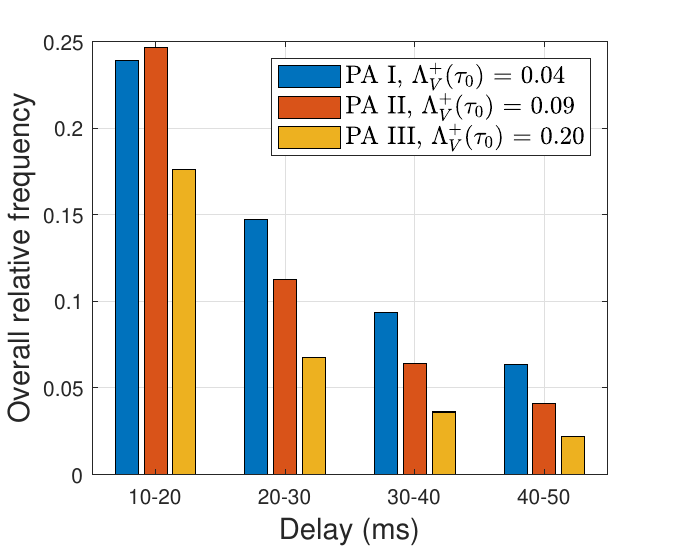}\label{v2v}}
 \hspace{1pt}
	\subfloat[]
 {\includegraphics[width=0.49\columnwidth]{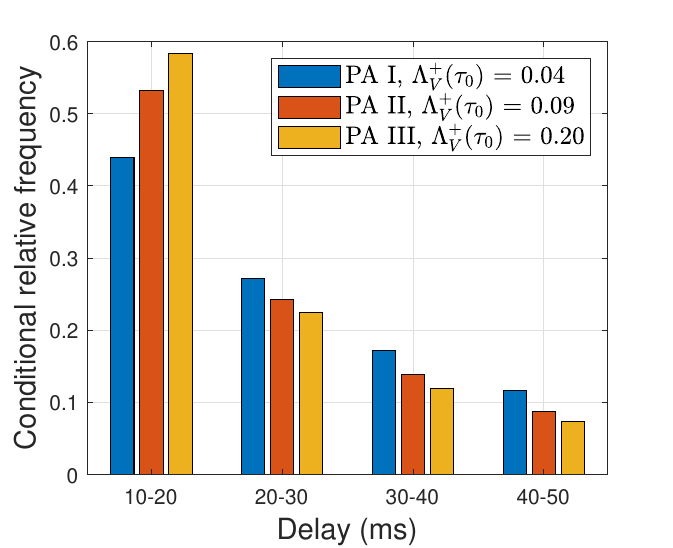}\label{v2v_c}}
	\caption{\small{\ac{QoS} of \ac{V2V} link during adaptation phase: a) \ac{ORF}, b) \ac{CRF}.}
     \vspace{-0.55cm}
 }
 \label{QoS_v}
   \vspace{-0.65cm}
\end{figure}
\begin{figure}[t]
\captionsetup[subfigure]{font=footnotesize}  
	\centering
	\subfloat[]
 {\includegraphics[width=0.49\columnwidth]{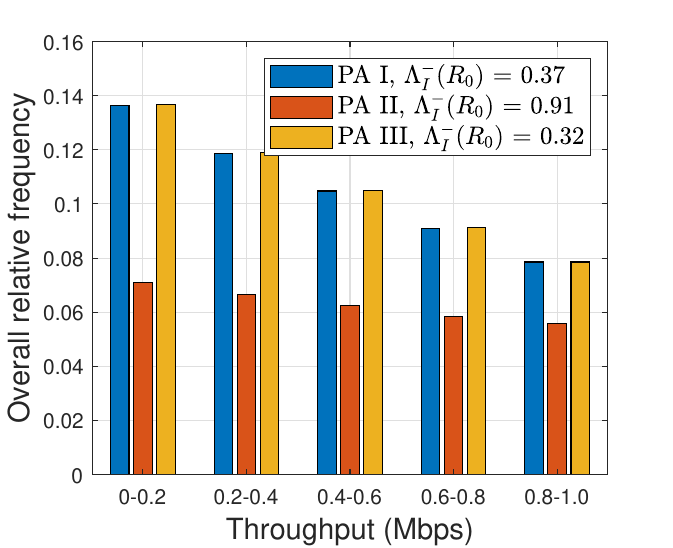}\label{v2i}}
 \hspace{1pt}
	\subfloat[]
 {\includegraphics[width=0.49\columnwidth]{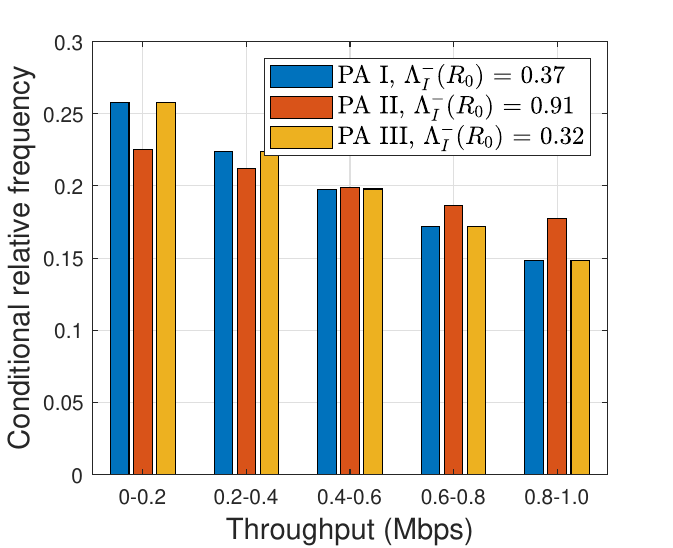}\label{v2i_c}}
	\caption{\small{\ac{QoS} of \ac{V2I} link during adaptation phase: a) \ac{ORF}, b) \ac{CRF}.}
      \vspace{-0.55cm}
 }
 \label{QoS_i}
   \vspace{-0.65cm}
\end{figure}

Figs. \ref{QoS_v} and \ref{QoS_i} show the impact of adaptation power scheme on the \ac{C-V2X}'s adaptation performance. The \ac{ORF} and \ac{CRF} of the \ac{QoS} violations during adaptation are presented. The \ac{ORF} is the ratio of the frequency of slots where \ac{QoS} falls within a specific interval to the adaptation phase length $T$. The \ac{CRF} is the ratio of the frequency of slots where \ac{QoS} falls within the interval to the frequency of slots where \ac{QoS} requirements are not met. In brief, the \ac{ORF} indicates the system's ability to meet \ac{QoS} requirements, while the \ac{CRF} reflects its capability to preserve the \ac{QoS} within a range (close to the \ac{QoS} requirements, from a resilience perspective). The \acp{HR} shown in Figs. \ref{QoS_v} and \ref{QoS_i} are given based on \eqref{HR v2v} and \eqref{HR v2i}. 

In Fig. \ref{v2v}, we can observe that the \ac{ORF} is lower under PA III compared to PA I and PA II, for all cases of delay requirement violation. Moreover, as shown in Fig. \ref{v2v_c}, there are fewer delay violation cases under PA III. Specifically, the \ac{CRF} of the delay within the range $[10,20]$~ms under PA III is $34 \%$ higher than the one achieved by PA I. Meanwhile, for PA III, the \ac{CRF} of the delay within the ranges of $[20,30]$~ms, $[30,40]$~ms, and $[40,50]$~ms is lower than the \ac{CRF} achieved by PA I and PA II. This is due to the fact that PA III achieves the highest \ac{HR} $\Lambda^{+}_V(\tau_0)$. Another interesting observation is that PA II leads to a higher \ac{HR} than PA I. This, in turn, indicates better system adaptation under PA II. However, the remediation capability under PA I and PA II is identical, as shown in Fig. \ref{pdf}. Thus, there is a need to design the adaptation powers $P_n^{\textrm{a}}$ and $P_m^{\textrm{a}}$ in a way to balance the tradeoff between adaptation performance and remediation capability. From Fig. \ref{QoS_i}, we also observe that the adaptation performance on the \ac{V2I} link under PA II is better than the one achieved by PA I and PA III. Particularly, the \ac{ORF} of all the throughput requirement violation cases is much lower, as shown in Fig. \ref{v2i}. Moreover, Fig. \ref{v2i_c} shows that, for PA II, the \ac{CRF} of the throughput within the ranges of $[0.6, 0.8]$~Mbps and $[0.8, 1.0]$~Mbps is higher than the corresponding \ac{CRF} resulting from PA I and PA III. This indicates that the throughput requirements are generally slightly compromised. 

This suggests that the throughput requirements are slightly compromised under PA II in these ranges.

In other words, even if the throughput requirements are not met, the cases of extremely low throughput under PA II are less frequent. Finally, we can observe that PA II outperforms PA I in both remediation capability and adaptation performance, according to Figs. \ref{pdf}, \ref{QoS_v}, and \ref{QoS_i}. Thus, the \ac{C-V2X} implementing PA II is more resilient to the disruption of imperfect \ac{CSI}, validating the effectiveness of the proposed framework to quantify and enhance system's resilience.

\vspace{-8pt}
\section{Conclusion}
\vspace{-6pt}
In this paper, we have proposed a novel analytical framework for examining the resilience of \ac{C-V2X} network in face of imperfect \ac{CSI}. Specifically, We have considered the adaptation phase of the \ac{C-V2X} network where the \ac{PDF} of the imperfection in \ac{CSI} is estimated. From a resilience perspective, we have then defined the remediation capability and adaptation performance of the \ac{C-V2X} network. Specifically, we have defined the \ac{MSE} of imperfect \ac{CSI}'s \ac{PDF} as the remediation capability and derived an explicit upper bound on the \ac{MSE}. Due to the tradeoff between the remediation capability and the \ac{C-V2X}'s \ac{QoS} during adaptation, we have introduced a novel metric named \ac{HR} to evaluate the \ac{C-V2X}'s adaptation performance. Finally, we have validated the framework's superiority in capturing the interplay between adaptation and remediation, as well as the effectiveness of the two proposed metrics in guiding the design of adaptation power scheme to enhance system's resilience.

    \vspace{-10pt}
\appendices
\section{Proof of Theorem 1}
    \vspace{-4pt}
\begin{proof}
First we can rewrite $\mathbb{E}\left[  (f_{E} - \hat{f}_{E})^2 \right] $ as following
\vspace{-7pt}
\begin{equation}
\label{A 1}
     \mathbb{E}\left[  (f_{E} - \hat{f}_{E})^2 \right] = \left[ f_{E} - \mathbb{E}\left( \hat{f}_{E} \right) \right]^2 + \mathbb{E} \left\{ \left[ \hat{f}_{E} - \mathbb{E}\left( \hat{f}_{E} \right) \right]^2 \right\}.
     \vspace{-5pt}
\end{equation}
To derive the first term $\left[ f_{E} - \mathbb{E}\left( \hat{f}_{E} \right) \right]^2$, we know $f_E =  \frac{1}{2 \pi} \int_{-\infty}^{\infty} e^{jwe} F\left\{ f_E \right\} dw $ by the inverse Fourier transform.  Based on  \eqref{hat pdf}, we can derive $\mathbb{E}\left( \hat{f}_{E} \right)$ as follows:
\vspace{-10pt}
\begin{equation}
    \begin{aligned}
        \mathbb{E}\left( \hat{f}_{E} \right)
        = & \frac{1}{2 \pi T} \int_{-\infty}^{\infty} \sum_t^T \int_{-K\pi}^{K\pi}  e^{-jw(z-e_{nm})}  (1+\frac{jw}{\lambda_Y}) dw f_{Z} dz \vspace{-3pt}\\ 
        = & \frac{1}{2 \pi} \int_{-K\pi}^{K\pi}   e^{jwe_{nm}} F\left\{ f_E \right\}  dw.
    \end{aligned}
    \vspace{-5pt}
\end{equation}
As a result, we can obtain
\vspace{-7pt}
\begin{equation}
    f_{E} - \mathbb{E}\left[ \hat{f}_{E} \right] = \frac{1}{2\pi} \int_{w\geq |K\pi|} e^{jwe_{nm}} F\left\{ f_E \right\} dw.
    \vspace{-5pt}
\end{equation}
For the second term in  \eqref{A 1}, we have
\vspace{-7pt}
\begin{equation}
\begin{aligned}
    & \mathbb{E} \left\{ \left[ \hat{f}_{E} - \mathbb{E}\left( \hat{f}_{E} \right) \right]^2 \right\} 
    = \mathbb{E} \left[ (\hat{f}_{E})^2 \right] - \mathbb{E}^2 \left (\hat{f}_{E} \right) \\
    \overset{(a)}{\leq} & \frac{1}{4 \pi^2 T}\mathbb{E} \left[ \left( \int_{-K \pi}^{K \pi}  e^{-jw(Z-e_{nm})}  (1+\frac{jw}{\lambda_Y}) dw\right)^2 \right] \\
    \overset{(b)}{\leq} &  \frac{1}{4 \pi^2 T}\mathbb{E} \left[ \left( \int_{-K \pi}^{K \pi}  \left\| e^{-jw(Z-e_{nm})}  (1+\frac{jw}{\lambda_Y}) \right\| dw \right)^2  \right] \\
    = & \frac{K^2}{4T}\left[ \sqrt{1+ o^2} + \frac{\ln{\left(o + \sqrt{1+ o^2}\right)}}{o} \right]^2,
\end{aligned}
    \vspace{-7pt}
\end{equation}
where in $(a)$ we remove the square term, and in $(b)$ we leverage Cauchy–Schwartz inequality and $o = K \pi (1-\delta_m^2) \frac{P^{\textrm{a}}_m L^{\textrm{a}}_m }{P^{\textrm{a}}_n L^{\textrm{a}}_n}$.

\end{proof}
\vspace{-15pt}

\bibliographystyle{IEEEtran}
\bibliography{bibliography}

\end{document}